\documentclass[letterpaper,twocolumn,10pt]{article}
\usepackage{usenix,epsfig,endnotes}
\usepackage{url}     
\usepackage{multirow}
\usepackage{stfloats}
\usepackage[numbers,sort&compress]{natbib}
\begin{document}

\date{}

\newcommand*{\affmark}[1][*]{\textsuperscript{#1}}

\title{\Large \bf Automated software vulnerability detection with machine learning}


\author{
  {\rm Jacob A. Harer\affmark[1,2], Louis Y. Kim\affmark[1], Rebecca L. Russell\affmark[1], Onur Ozdemir\affmark[1], Leonard R. Kosta\affmark[1,2]}, \\ {\rm Akshay Rangamani\affmark[3], Lei H. Hamilton\affmark[1], Gabriel I. Centeno\affmark[1,4], Jonathan R. Key\affmark[1],} \\   {\rm Paul M. Ellingwood\affmark[1], Erik Antelman\affmark[1], Alan Mackay\affmark[1], Marc W. McConley\affmark[1],} \\ {\rm Jeffrey M. Opper\affmark[1], Peter Chin\affmark[2], Tomo Lazovich\affmark[1]} \\
  \affmark[1] Draper\\
  \affmark[2] Boston University, Department of Computer Science\\
  \affmark[3] ECE Department, Johns Hopkins University\\
  \affmark[4] Northeastern University, College of Computer and Information Sciences \\
}

\maketitle

\thispagestyle{empty}

\subsection*{Abstract}
Thousands of security vulnerabilities are discovered in production software each year, either reported publicly to the Common Vulnerabilities and Exposures database or discovered internally in proprietary code. Vulnerabilities often manifest themselves in subtle ways that are not obvious to code reviewers or the developers themselves. With the wealth of open source code available for analysis, there is an opportunity to learn the patterns of bugs that can lead to security vulnerabilities directly from data. In this paper, we present a data-driven approach to vulnerability detection using machine learning, specifically applied to C and C++ programs. We first compile a large dataset of hundreds of thousands of open-source functions labeled with the outputs of a static analyzer. We then compare methods applied directly to source code with methods applied to artifacts extracted from the build process, finding that source-based models perform better. We also compare the application of deep neural network models with more traditional models such as random forests and find the best performance comes from combining features learned by deep models with tree-based models. Ultimately, our highest performing model achieves an area under the precision-recall curve of 0.49 and an area under the ROC curve of 0.87. 

\section{Introduction}
\label{sec:intro}

Hidden flaws in software often lead to security vulnerabilities, wherein attackers can compromise a program and force it to perform undesired behaviors, including crashing or exposure of sensitive user information. Thousands of such vulnerabilities are reported publicly to the Common Vulnerabilities and Exposures database (CVE) each year, and many more are discovered internally in proprietary code and patched~\cite{CVE,BugOccurrence}. As we have seen from many recent high profile exploits, including the Heartbleed bug and the hack of the Equifax credit history database, these security holes can often have disastrous effects, both financially and societally~\cite{yadron2014after,equifax}. These vulnerabilities are often due to errors made by programmers and can propagate quickly due to the prevalence of open source software and code re-use. While there are existing tools for static (pre-runtime) or dynamic (runtime) analysis of programs, these tools typically only detect a limited subset of possible errors based on pre-defined rules. With the recent widespread availability of open source repositories, it has become possible to use data-driven techniques for discovering the patterns of bug manifestation. In this paper, we present techniques using machine learning for automated detection of vulnerabilities in C and C++ software\footnote{While our work focuses on C/C++, the techniques are applicable to any language}. We focus on vulnerability detection at the individual function level (i.e. each function within a larger codebase is analyzed to determine whether or not it contains a vulnerability).

\subsection{Related work}
\label{sec:related_work}

There currently exists a wide range of static analysis tools which attempt to uncover common vulnerabilities. While there are too many to list in detail here, they range from open source implementations such as the Clang static analyzer~\cite{xu2010memory} to proprietary solutions~\cite{fortify}. In addition to these rule-based tools, there has been much work in the domain of using machine learning for program analysis. This work spans various application domains, from malware detection~\cite{schultz2001data,kolter2006learning,cesare2011malware,shabtai2009detection,abou2004n} to intelligent auto-completion of source code~\cite{raychev2014code,raychev2015predicting,hsiao2014using,white2015toward}. Additionally, they span levels of program representations, from pure source code, to abstract syntax trees, and finally compiled binaries~\cite{binaries,yamaguchi2012generalized}. For a comprehensive review of learning from ``big code", see the recent work by Allamanis et. al.~\cite{BigCodeReview}. There are also several existing tools for automated repair, which both find and attempt to fix potential bugs. The GenProg program uses genetic algorithms to generate repair candidates for a given error~\cite{le2012genprog}. The Prophet program builds a probabilistic model of potentially buggy code and then uses this model to generate repair candidates~\cite{long2016automatic}. Both of these methods require access to test cases that trigger the failure. The DeepFix algorithm from Gupta et. al. applies deep learning to generating fixes for simple syntax errors in student code~\cite{gupta2017deepfix}. Our unique contribution in this work is three-fold. First, we provide a systematic exploration of both feature-based and source-based vulnerability detection. Second, we explore the effectiveness of both traditional machine learning and deep learning methods. Finally, we do not limit ourselves to specific error types, instead training only on labels of buggy or not, and end up spanning over fifty different categories of weakness in the Common Weakness Enumeration (CWE)~\cite{cwe}.

\begin{figure*}[t!]
\centering{}
  \includegraphics[width=0.9\textwidth]{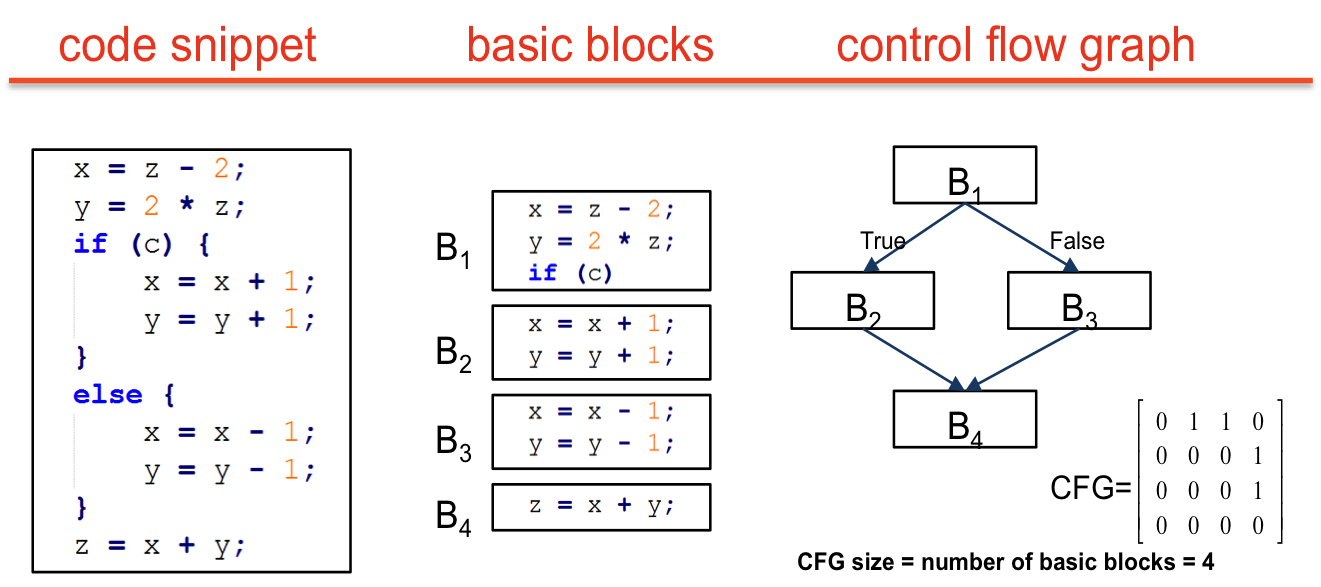}
  \caption{A control flow graph extracted from a simple example code snippet.}
  \label{fig:CFG_example}
\end{figure*}

\section{Methods}
\label{sec:classification}

In order to detect vulnerabilities in C and C++ code, we consider two complementary approaches. The first uses features extracted from the intermediate representation (IR) of programs produced during the compilation and build process. The second operates on source code directly. These two approaches offer independent sources of information. The source-based models can learn statistical correlations in how code is written, leveraging techniques from natural language processing, but don't have any inherent knowledge about the structure or semantics of the language built in (these must all be learned from data). The build-based models take advantage of features extracted from a compiler that understands the structure of the language, but these features may abstract away certain properties of the code that are useful for vulnerability detection. By pursuing both approaches, we allow the possibility of ultimately fusing multiple models to increase detection performance. In this section, we first define the features extracted for both the build-based and source-based features, and then discuss the machine learning models used for vulnerability detection.

\subsection{Build-based feature extraction}

\label{sec:build-features}
In build-based classification, the first step is to extract features from the source using Clang and LLVM tools~\cite{lattner2004llvm}. 
During the build process we extract features at two levels of granularity. At the function level, we extract the \textbf{control flow graph (CFG)} of the function. Within the control flow graph, we extract features about the operations happening in each basic block (\textbf{opcode vector, or op-vec}) and the definition and use of variables (\textbf{use-def matrix}). The CFG is a graph representation of the different paths a program can take during execution. Each node is a basic block - a unit of code with no branching behavior. The edges of the graph connect basic blocks that can flow into each other during execution and occur at control flow operations. At a high-level, this representation is useful for vulnerability detection because models can potentially learn program execution topologies that are risky. Figure~\ref{fig:CFG_example} shows an example code snippet and the resulting CFG. \\\\
The high-level view of the CFG is complemented by features extracted from each basic block within the CFG, so that models can also learn instruction-level behaviors that are associated with vulnerable code. Within each basic block, we extract features which define the behavior of that basic block. The first of these features 
is the use-def matrix. This matrix tracks, within the basic block, the locations of instructions where variables are defined (def) and used (use). If a variable is defined at instruction $i$ and used in instruction $j$, then both the $(j,i)$ entry of the use-def matrix are set to $1$. The second feature extracted for each basic block is the op-vec, or opcode vector. LLVM assigns opcodes to instructions in one of nine different categories: conditional, arrogate, binary, bit binary, conversion, memory address, termination, vector operation, and other. The opcode vector for a basic block is a binary vector with a single entry for each of these possible classifications. If the basic block contains an instruction with the corresponding opcode, then that bit is set to $1$ for the basic block.

\subsection{Source-based feature extraction}
\label{sec:source-features}
\begin{figure}[t!]
\centering{}
  \includegraphics[width=0.45\textwidth]{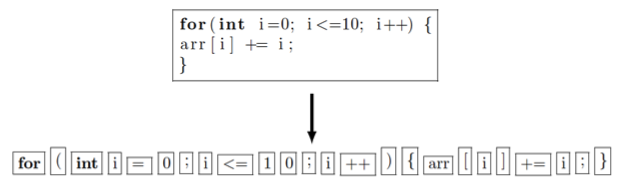}
  \caption{An example illustrating the lexing process.}
  \label{fig:Lexer_example}
\end{figure}

To feed the source-based classifier models, we need a feature representation of the individual tokens that occur in source code. We implement a custom C/C++ lexer, which parses the code and categorizes elements into different bins: comments (which are currently removed), string literals, single character literals, multiple character literals, numbers, operators, pre-compiler directives, and names. Names are further sub-categorized into keywords (such as \texttt{for}, \texttt{if}, etc.), system function calls (such as \texttt{malloc}), types (such as \texttt{bool}, \texttt{int}, etc.), or variable names. Variable names are all mapped to the same generic identifier, but each unique variable name within a single function gets a separate index (for the purposes of keeping track of where variables re-appear), following the procedure of Gupta et. al.~\cite{gupta2017deepfix}. This mapping changes from function-to-function based on the position in which the identifier first appears. Unique string literals, on the other hand, are all mapped to the same single identifier. Finally, integer literals are read digit-by-digit, while float literals are mapped to a common token. Figure~\ref{fig:Lexer_example} shows an example of the lexing process.

\begin{figure*}[t]
  \centering{}
  \includegraphics[width=0.99\textwidth]{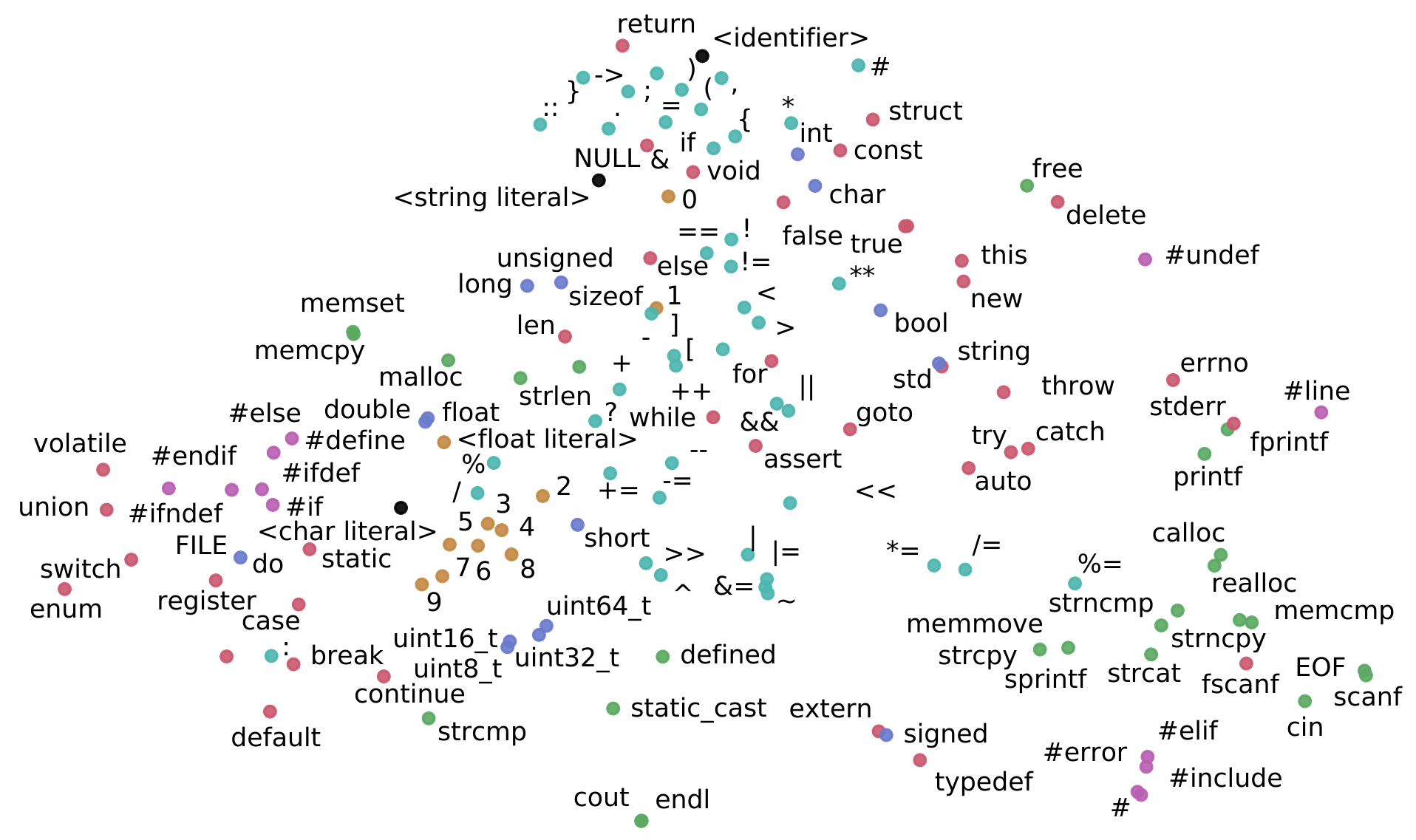}
  \caption{A word2vec embedding of tokens from C/C++ source code}
  \label{fig:word2vec}
\end{figure*}

Once the tokens are lexed, the sequence of tokens can be converted into a vectorial representation. The first representation we extract is the commonly used bag-of-words vector, where a document is represented by the frequency with which each token in the vocabulary appears~\cite{manning1999foundations}. The second representation we use is learned with the word2vec algorithm~\cite{word2vec}. In this model, word vectors are representations of tokens that are useful for predicting the surrounding context in which they appear. Both of these representations are unsupervised, meaning they do not require labeled examples in order to be used. Figure~\ref{fig:word2vec} shows a two-dimensional representation of a word2vec representation learned from 6.3 million examples of C and C++ functions. Notice that similar types of tokens cluster together in the word2vec representation, giving us confidence that this feature representation appropriately captures token context. 

\subsection{Labels}
\label{sec:labels}
Labeling at the function level in order to train a machine learning model is a significant challenge. We do not know of a source of C/C++ functions with ground truth labels that has the volume required to train most machine learning models and also has the complexity and diversity of real-world code. As a result, we generate labels for each function during the feature extraction process. To generate the labels, we run the Clang static analyzer (SA) on each source repository~\cite{ClangSA}. After pruning the Clang outputs to exclude warnings that are not typically associated with security vulnerabilities, we match the findings to the functions in which they occurred. The labels are then binarized, meaning functions with no SA findings are labeled as ``good'' and functions with at least one SA finding are labeled as ``buggy"\footnote{Future work could include using the type of SA finding to do bug classification rather than just detection}. In our convention, good is the negative label and buggy is the positive label. This means that if our detection model labels a function as buggy and there was also a static analysis finding, this would be counted as a true positive. Similarly, labeling a function as buggy when it does not have a static analysis finding is counted as a false positive. While in this work we have presented a comparison to static analysis labels as a way of benchmarking performance, the ultimate system we hope to provide will go beyond static analysis label prediction. By incorporating other sources of data such as the Juliet test suite~\cite{juliet} and bug report repositories, we can expand the label set used to train models. However, in this work, we compare solely to SA labels to evaluate performance. 

\subsection{Models}
\label{sec:models}

In constructing machine learning models for detecting buggy functions, we consider two objectives. First, we aim to evaluate the overall possible performance - how well each model can predict the presence of unsafe practices in the code. Second, we aim to compare the performance of the build-based and source-based features to discover which set is more predictive of code quality. Therefore, we construct separate source-based and build-based models and also consider a combined approach. 

\begin{figure*}[h]
  \centering{}
  \includegraphics[width=0.9\textwidth]{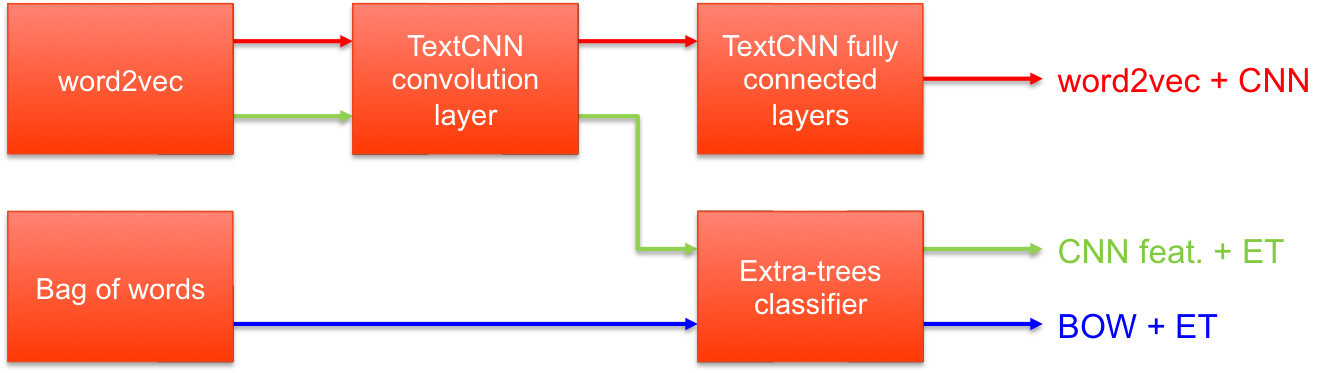}
  \caption{Illustration of the three types of source-based models tested}
  \label{fig:source-options}
\end{figure*}

\subsubsection{Source-based models}
\label{sec:source-options}
Our source-based models take in a feature vector derived directly from a single function's source code and output a score corresponding to the likelihood that the given function contains a bug. Because we have multiple possible feature vector representations of the source code (both the bag-of-words and a sequence of word2vec vectors), we build multiple models. For a baseline traditional model, we choose the extremely random trees (extra-trees) classifier\footnote{This was found to perform better than other traditional models such as random forest and SVM on our validation set} which takes the bag-of-words vector as input~\cite{extratrees}. As an alternative, we also consider an enhanced version of the TextCNN model, a convolutional neural network originally developed for sentence classification~\cite{TextCNN}. This model begins with a feature embedding layer that is initialized with the learned word2vec representation of section~\ref{sec:source-features}. This representation is allowed to be fine-tuned during training. The sequence of feature vectors is then passed through a convolutional layer which feeds into multiple fully connected layers. Finally, the probability that the function is buggy is output. The final model we consider is one where the features learned by the convolutional layer of the TextCNN are passed into the extra-trees classifier rather than the fully connected layer of the neural network. Figure~\ref{fig:source-options} illustrates the three different model options we consider.

\subsubsection{Build-based models}

Our build-based models use the features extracted from the build process described in section~\ref{sec:build-features}. One of the major challenges with using the build-based features is the construction of a feature vector from the CFG adjacency matrix and op-vec/use-def vectors as these are at different granularities (CFG at the function level, others at the basic block level) and very in size with the size of the CFG and basic blocks respectively. To deal with this variation, we construct a hand-crafted fixed-size vector from these features. In this approach, we simply average over the number of basic blocks to create an average use-def matrix and op-vec for the function. We combine the average upper triangular of the use-def matrix truncated to size $15\times15$ ($105$ entries), the $9$ possible opcode classifications, the average of the CFG's adjacency matrix (a measure of connectedness), and the number of basic blocks. This results in a $N=116$ dimensional feature vector. This vector can be fed into any machine learning model, and for this study we chose a \textit{random forest}.

In addition to comparing the results of the source-based and build-based models, we also produce a combined model which uses a concatenation of the build-based vector and the source-based bag of words vector.

\section{Experimental results}
\label{sec:results}

In this section, we present the results of training and evaluating the different classifiers described in section~\ref{sec:models}. We use two different datasets of open source C/C++ code to do the evaluation, and split the data into three parts: training, validation, and test. The training set is used to perform supervised learning and train the models to predict the bugginess of a function. The validation set is used for tuning hyperparameters of each model. The test set is used only for performance evaluation. All the results shown in this section are evaluation on the test set.  

\subsection{Datasets}
\label{sec:datasets}





\begin{table}[h]
\centering{}
\begin{tabular}{|c|c||c|c|c|}

\hline
&  & \textbf{Debian} & \textbf{Github} & \textbf{Source} \\ \hline

\multirow{2}{*}{\textbf{Train}} & Good & 429807 & 250436 & 754150 \\ \cline{2-5}
& Buggy & 14616 & 9556 & 31790 \\ \hline

\multirow{2}{*}{\textbf{Valid.}} & Good & 66100 & 38500 & 94268 \\ \cline{2-5}
& Buggy & 1878 & 1220 & 3724 \\ \hline

\multirow{2}{*}{\textbf{Test}} & Good & 55102 & 32109 & 94268 \\ \cline{2-5}
& Buggy & 1857 & 1203 & 3724 \\ \hline

\end{tabular}

\caption{Number of functions in both datasets used for model evaluation. Debian and Github represent the separated build-based datasets, while Source represents the dataset of all source code combined from both datasets.}
\label{tab:stats}
\end{table}

We evaluate the performance of models on two distinct datasets of open source C/C++ code. The first is the full set of C/C++ packages distributed with the Debian Linux distribution~\cite{debian}. The second is a large set of C/C++ functions pulled from public Git repositories on Github~\cite{github}. These two datasets give unique benefits. The Debian packages and releases are very well-managed and curated, meaning that this dataset ideally consists of relatively stable code. This code is still susceptible to vulnerabilities, but it is at least in use on many systems today. The Github dataset provides a wider variety of code, both in function and in quality. Table~\ref{tab:stats} shows the statistics of each dataset, including the number of good and buggy functions and the split between train, validation, and test. The source-based models have more data available because the code is not required to have successfully built. We construct one combined source-based dataset with all of the lexed functions available in the Debian and Github datasets. For the build-based models, we construct separate datasets for each source. This allows us to compare the performance on different code samples while also comparing the performance between the build-based and source-based models. To increase the amount of source code available for training, we process multiple Debian releases and multiple revisions of source repositories in Github\footnote{As a result, the statistics in table~\ref{tab:stats} should not be taken as the number of functions or static analysis findings in the current Debian release, but rather an aggregation over many releases}. 

One very important aspect of splitting the dataset is also the removal of potential duplicate functions. When pulling open source repositories there is the possibility of functions being duplicated across different packages. Such duplication, when not properly accounted for, can artifically inflate performance metrics if functions that appeared in the training set also appear in validation or test sets. In order to account for this, we perform a duplicate removal routine. For the source-based dataset, any function that has the same lexed representation as another function (using anonymous variables) is removed from the dataset. In the build-based dataset, any function with a duplicated lexed representation \textit{or} a duplicate build feature vector is removed. This is the \textit{strictest possible} duplicate removal and thus gives the \textit{most conservative} performance results. In practice, when applying this tool to new packages in the wild, it is indeed possible that the model would come upon a function it had already seen in the training set. However, estimating this duplication probability is difficult and thus we choose the most conservative approach for evaluating our models. 

\subsection{Source-based detection}
\label{sec:source-based}

We compare the results of the three different source-based models described in section~\ref{sec:source-options}. To evaluate them, we compute both the receiver operating characteristic (ROC) curve and the precision-recall curve for their outputs. These are constructed by choosing a score threshold and then computing the false positive rate and true positive rate (in the case of the ROC curve) or the precision and recall. Thus, each curve is formed by scanning through different score thresholds. As a reminder, the labels derived from a static analysis tool as described in section~\ref{sec:labels} are treated as ground truth for the purpose of performance benchmarking. Table~\ref{tab:source-results} shows the area under the curve for both metrics in each of the three models tested. 

\begin{table}
\centering
\begin{tabular}{|c|c|c|}
\hline
Model & ROC AUC & P-R AUC \\ \hline
BoW + ET & 0.85 & 0.44 \\ \hline
word2vec + CNN & \textbf{0.87} & 0.45 \\ \hline
CNN feat. + ET & \textbf{0.87} & \textbf{0.49} \\ \hline
\end{tabular}
\caption{Summary statistics for the performance of different source-based classifier options}
\label{tab:source-results}
\end{table}

The best performance comes from a model using features learned by the TextCNN as inputs into an extra-trees classifier. This gives an ~10\% improvement in area under the precision-recall (P-R) curve over the standard TextCNN initialized with a pre-trained word2vec input. However, both models perform similarly in the ROC AUC metric. In this problem setting, the precision and recall are more relevant metrics due to the severe label imbalance. ROC curves are not sensitive to label imbalance while precision and recall can vary greatly depending on how the labels are balanced. 

\begin{figure*}[h]
  \centering{}
  \includegraphics[width=0.49\textwidth]{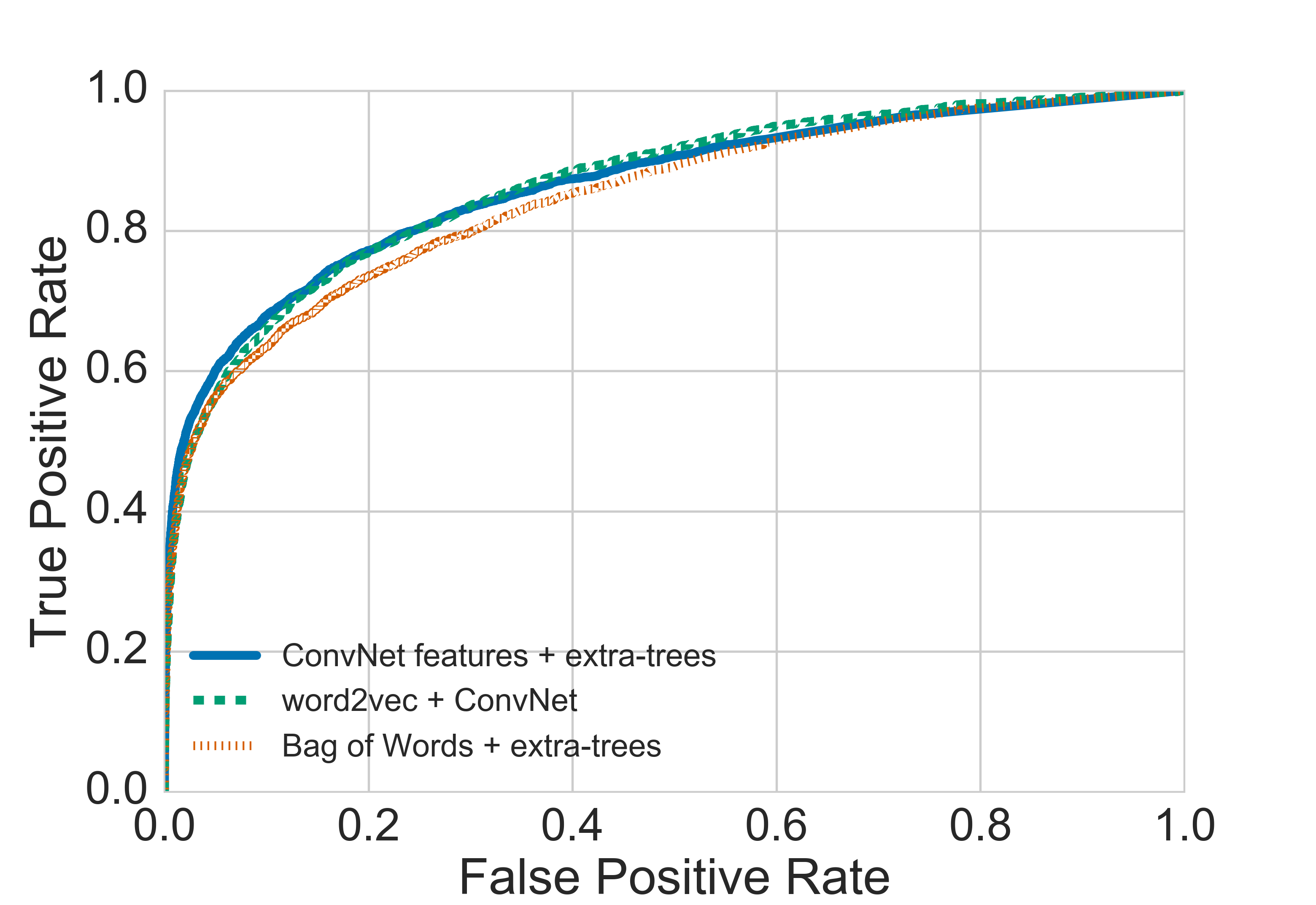}
  \includegraphics[width=0.49\textwidth]{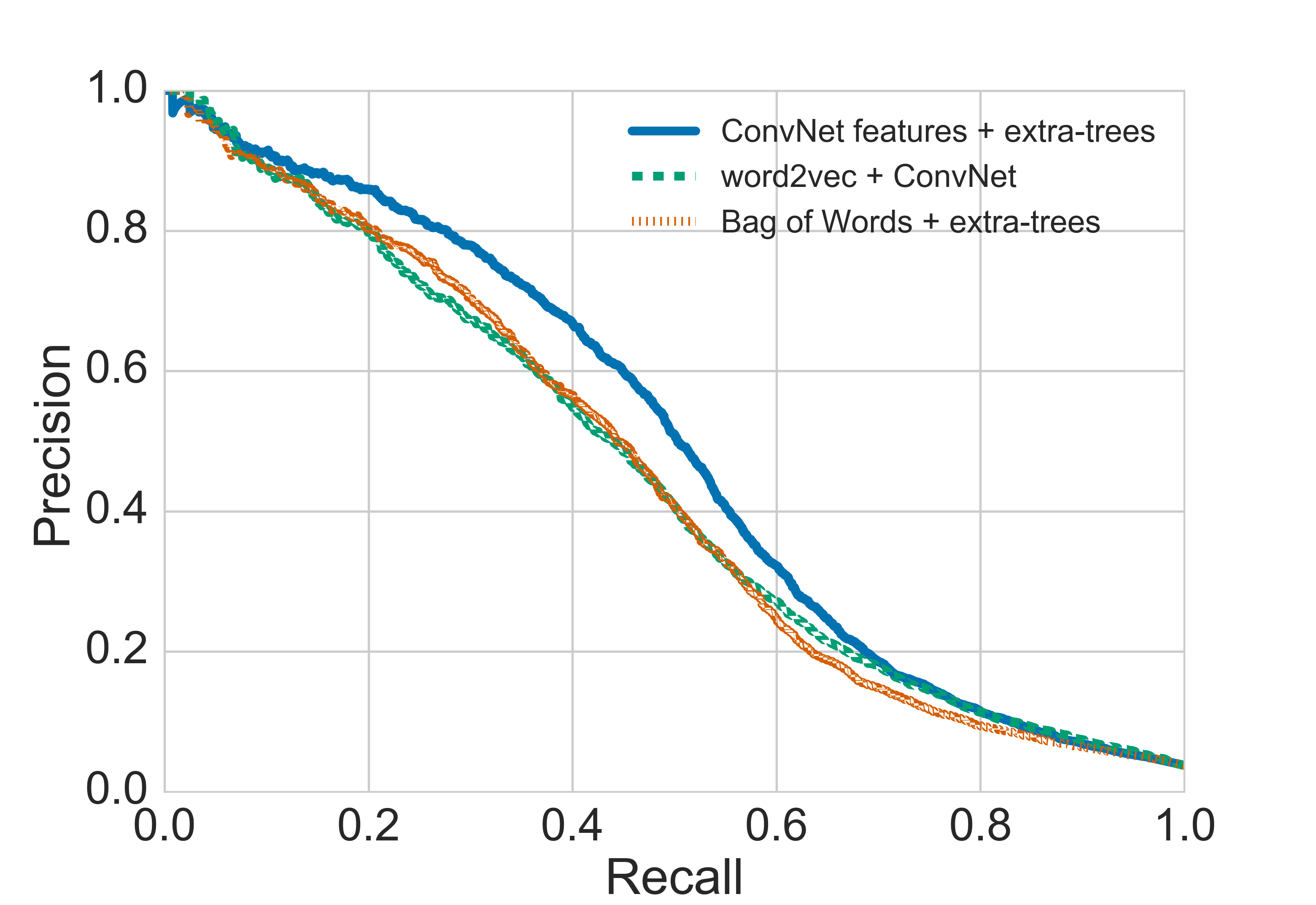}
  \caption{Results of testing different source-based models on a combined Debian+Github source code dataset. The left shows ROC curve results while the right shows the precision-recall curve.}
  \label{fig:Source_Results}
\end{figure*}

\subsection{Build-based detection}
\label{sec:build-based}

Here, we compare the performance of the build-based classifier on the Debian and Github datasets. The results for the ROC and P-R AUC metrics can be seen in table~\ref{tab:build-results}. We note two trends here. First, we see that the build-based classifiers generally perform worse than the source-based classifiers. It should be noted that this dataset has a lower number of functions available than the full combined source dataset, as can be seen in table~\ref{tab:stats}. In section~\ref{sec:combined}, we perform a comparison of thw two techniques on the same dataset. Second, we see that model performance is very similar on the two different datasets. This suggests that the model can learn about different types of code, and future work will test generalization across datasets. 

\begin{table}
\centering
\begin{tabular}{|c|c|c|}
\hline
Dataset & ROC AUC & P-R AUC \\ \hline
Debian & 0.76 & 0.21 \\ \hline
Github & 0.74 & 0.22 \\ \hline
\end{tabular}
\caption{Summary statistics for the performance of the build-based classifier on both the Debian and Github datasets}
\label{tab:build-results}
\end{table}

\subsection{Comparison and combined model performance}
\label{sec:combined}
As a final test, we compare the performance of the source and build-based models on a dataset consisting of the same set of functions. While this does not encompass the full set of data available to the source-based classifier, it allows us to perform an apples-to-apples comparison of performance. We re-train a source-based model on the Github dataset alone and compare the results to the build-based model on that dataset. We also train a combined model which takes both the build-based feature vector and the bag-of-words source vector. The results of these comparisons are shown in table~\ref{tab:comparison}.

\begin{table}
\centering
\begin{tabular}{|c|c|c|}
\hline
Model & ROC AUC & P-R AUC \\ \hline
Build & 0.74 & 0.22 \\ \hline
Source & 0.81 & 0.29 \\ \hline
Combined & \textbf{0.82} & \textbf{0.32} \\ \hline
\end{tabular}
\caption{Comparison of build only, source only, and combined models on the Github dataset.}
\label{tab:comparison}
\end{table}

\begin{figure*}[t]
  \centering{}
  \includegraphics[width=0.49\textwidth]{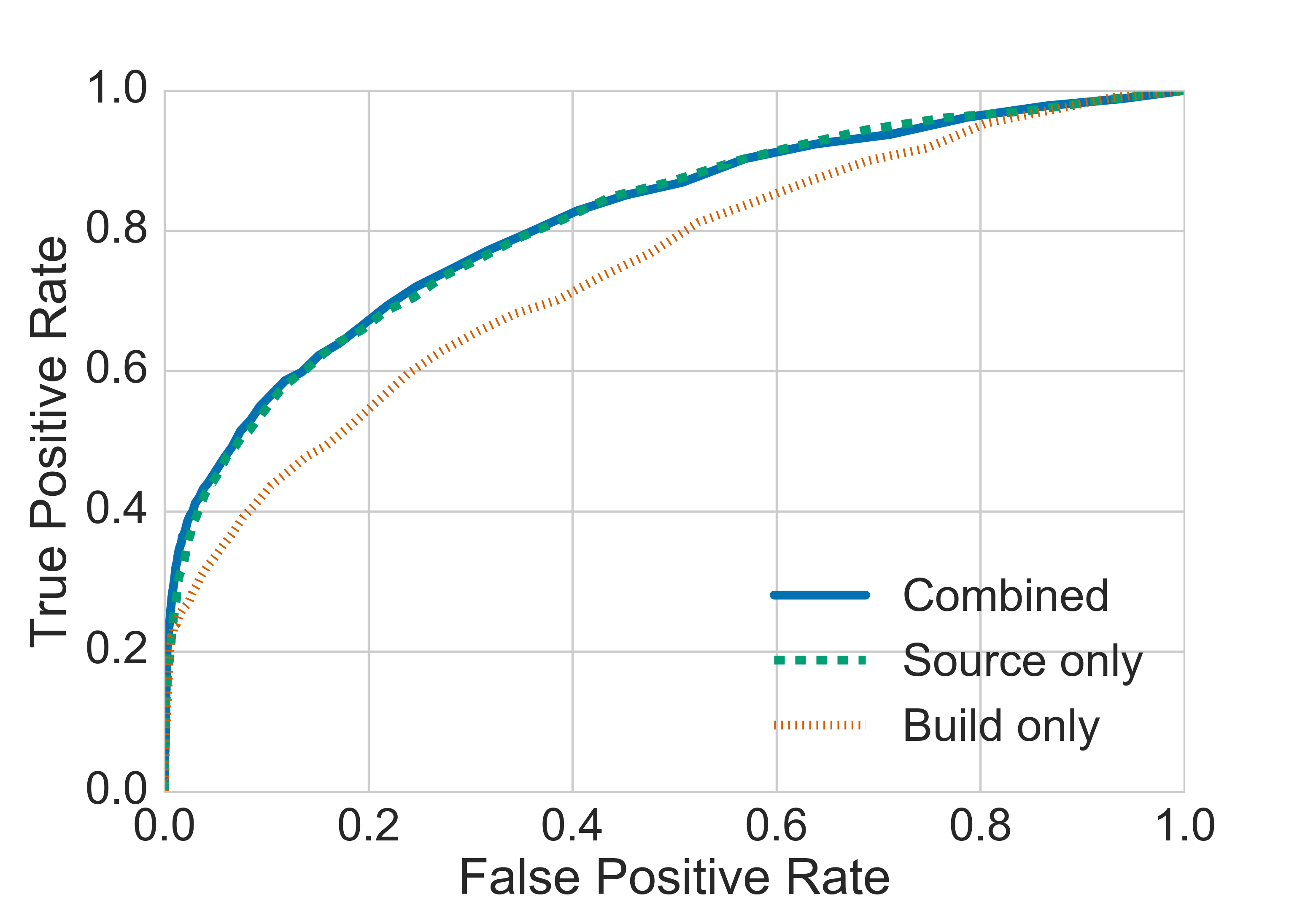}
  \includegraphics[width=0.49\textwidth]{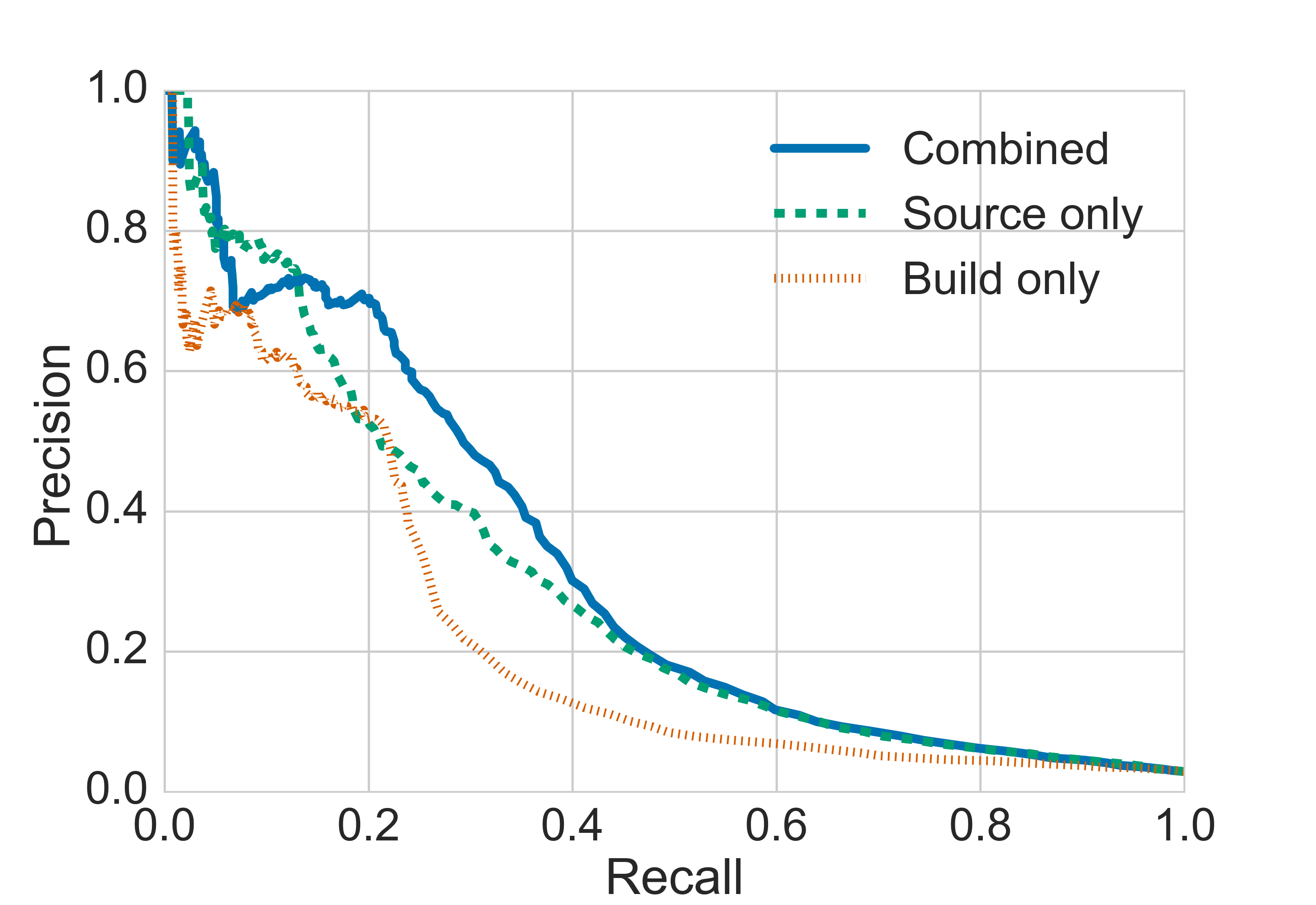}
  \caption{Comparison of build only, source only, and combined models on the Github dataset.}
  \label{fig:Combined_Results}
\end{figure*}

Even when comparing on the same dataset, source-based approaches offer an advantage over the build-based approaches, for both ROC AUC and in the precision-recall space. We also illustrate that the build-based features do add additional information that the source code itself does not provide, as the combined model performs better than either model does individually. Figure~\ref{fig:Combined_Results} shows the comparison of both metric curves. 

\section{Conclusions and future work}
\label{sec:conclusions}

In this work, we have implemented machine learning models for the detection of bugs that can lead to security vulnerabilities in C/C++ code. Using features derived from the build process and the source code itself, we have shown that a variety of machine learning techniques can be effective at predicting the output of static analysis tools at the function level. The main limitation of this work is in the labeling of the functions. While we have treated the static analysis output as ground truth in this study to illustrate the effectiveness of the machine learning approach, ultimately the techniques should be trained on additional sources of labels to be able to function robustly in a real-world setting. This could come from a fusion of multiple static analysis tools or from the curation of a dedicated, large set of example programs with known ground truth. Ultimately, we believe that the data-driven approach can complement existing static analysis tools and decrease the amount of time it takes to discover potential vulnerabilities in software. By ranking functions by their probability of bugginess, developers can more rapidly complete code reviews and isolate potential issues that may not have been uncovered otherwise. 

\subsection*{Acknowledgments}

The authors thank Thomas Jost, Mark Lutian, Hugh J. Enxing, Nicolas Edwards, and Adam Zakaria for their heroic efforts creating the build and data ingestion pipeline that was used to generate the datasets for this work. This project was sponsored by the Air Force Research Laboratory (AFRL) as part of the DARPA MUSE program. 


{\footnotesize \bibliographystyle{acm}
\bibliography{references}

\end{document}